\documentclass[11pt,twoside]{article}

\usepackage{asp2006}
\usepackage{epsf}
\usepackage{psfig}
\usepackage{lscape}

\begin{document}
\title{V2491 Cyg - a possible recurrent nova ?}   
\author{E. Ragan$^{1}$ , M. Miko\l ajewski$^{1}$, T. $\rm Tomov^{1}$, E. Swierczy\'{n}ski$^{1}$, T. Bro\.{z}ek$^{1}$, C. Ga\l an$^{1}$, P. R\'{o}\.{z}a\'nski$^{1}$, M. Wi\c{e}cek$^{1}$, P. Wychudzki$^{1,2}$}   
\affil{$^{1}$ Centre for Astronomy of Nicolaus Copernicus University, Toru\'n, Poland\\
$^{2}$ Olsztyn Planetarium and Astronomical Observatory, Olsztyn, Poland}    

\begin{abstract} 
Nova V2491 Cyg was discovered on April 10.72 UT 2008 \citep{Nak08}. Here we present spectrophotometric premises that V2491 Cyg can be a good candidate for recurrent nova (RNe). Its properties are compared to five well known RNe with red dwarf secondaries (U Sco, V394 Cra, T Pyx, CI Aql, IM Nor) and recently confirmed as recurrent nova V2487 Oph \citep{Pag08}. Photometric $U, B, V, R_C, I_C$ and moderate resolution (R$\sim 1500$) spectral observations of V2491~Cyg were carried out in the Torun Observatory (Poland) between April 14 and May 20 2008.
\end{abstract}


\section{Optical spectrum of V2491 Cyg, mass and magnetic field of the white dwarf}  {In the spectrum of V2491~Cyg obtained at $\Delta t = +3.7^{d}$ the emmision lines NII~5679\AA\ and NIII~4640\AA\, typical for  He/N novae, are apparent. Remarkable emission features mainly of HeII, NII and NIII are presented as well. The strong OI 8446 line suggests a \citet{wil92} $P_{n}^{o}$ type. Several days later at $\Delta t = +12.7^{d}$ the blend NIII, HeII at $\sim$4650\AA\ is stronger than H$\beta$ and its width implies probable presence of CIII as observed in other RNe \citep{Mu99,Du03,Se89}.  Another similarity between the mentioned above RNe and V2491~Cyg is the extreme width of the emission lines during the early outburst phase (left panel Figure 1.).

V2491 Cyg is the second nova observed in X-rays before the outburst. The first one was the newest member of the RNe group V2487 Oph. On the base of the X-ray observations a magnetic white dwarf primary was suggested for the two systems \citep{Ib08,He02,Ta09}. Both hard and soft X-ray behavior of V2491~Cyg  was similar to that observed in the recurrent nova RS Oph which consists of a massive white dwarf and a red giant. Basing of this, \citet{Pa08} and \citet{Os08} suggested that the relatively early emergence of the super-soft phase may indicate the presence of a massive white dwarf in V2491~Cyg as well. \citet{Ha09} estimated the mass of the white dwarf in V2491~Cyg to be $1.3\rm{M}\odot$ and assumed that the system is a polar with a $10^{7}$G magnetic field. Such a mass is in good agreement with the masses derived for most RNe which are between $1.2\rm{M}\odot$ and $1.4\rm{M}\odot$ (see Hachisu \& Kato 2001 and references therein).
\begin{figure}[ht]
\plotone{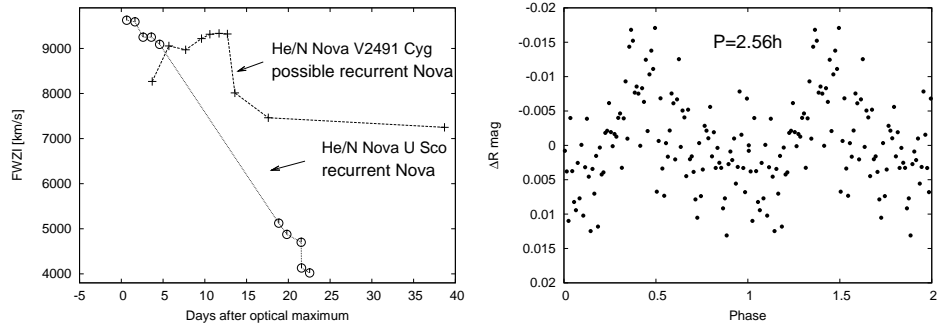}
\caption{Left panel: Comparison of variations of the H$\alpha$ FWZI in the postmaximum spectra of V2491~Cyg, and U~Sco (the data for U Sco is extracted from Figure 4 in Munari et al. (1999)). Right panel: R band light curve of V2491 Cyg observed on 9, 11 and 15 May 2008 and phased with $P=0.1067^{d}$. Each point represents an average value binned in 90 seconds intervals.}
\end{figure}

\section{Outburst amplitude and orbital period of V2491 Cyg}
Most of RNe have relatively smaller outburst amplitudes in comparison to the classical novae. On the base of \citet{Ri03} data we estimated the outburst amplitude as $10^{m}$ which is located in the range of values obtained for other RNe. 

Fourier analysis of our R band observations revealed a period of $0.1067^{d}$, very close to the period ($0.0958^{d}$) and its one-day alias ($0.10595^{d}$) \citet{Ba08} (right panel in Figure 1). Such a short period puts V2491~Cyg together with IM Nor in the middle of the ``period gap'' of cataclasmic variables. 

 
 
 
 
 
 

\acknowledgements This work was supported by Polish MNiSW Grant N203 018 32/2338, UMK grants no. 366-A, no. 367-A, and from resources of European Social Fund and Polish Government within Integrated Regional Development Operational Programme, Action 2.6, by project "PhD fellowships 2008/2009 - ZPORR" of Kuyavian-Pomeranian Voivodeship.


\end{document}